\documentclass[aps,prd,preprint,groupedaddress]{revtex4}

\usepackage{epsfig}

\begin{document}

\def\lp{\left. }
\def\rp{\right. }
\def\lr{\left( }
\def\rr{\right) }
\def\le{\left[ }
\def\re{\right] }
\def\lg{\left\{ }
\def\rg{\right\} }
\def\lb{\left| }
\def\rb{\right| }

\def\go{\tilde{g}}
\def\mg{m_{\go}}

\def\tg{t_{\go}}
\def\ug{u_{\go}}

\def\ts{t_{\tilde{q}}}
\def\us{u_{\tilde{q}}}

\def\msQ {m_{\tilde{Q}}}
\def\msD {m_{\tilde{D}}}
\def\msU {m_{\tilde{U}}}
\def\msS {m_{\tilde{S}}}
\def\msC {m_{\tilde{C}}}
\def\msB {m_{\tilde{B}}}
\def\msT {m_{\tilde{T}}}
\def\msusy{m_{\rm SUSY}}

\def\sq  {\tilde{q}}
\def\sql {\tilde{q}_L}
\def\sqr {\tilde{q}_R}
\def\ms  {m_{\sq}}
\def\msql{m_{\tilde{q}_L}}
\def\msqr{m_{\tilde{q}_R}}

\def\st  {\tilde{t}}
\def\stl {\tilde{t}_L}
\def\str {\tilde{t}_R}
\def\mstl{m_{\stl}}
\def\mstr{m_{\str}}
\def\sta {\tilde{t}_1}
\def\stb {\tilde{t}_2}
\def\msta{m_{\sta}}
\def\mstb{m_{\stb}}
\def\thst{\theta_{\tilde{t}}}

\def\sb  {\tilde{b}}
\def\sbl {\tilde{b}_L}
\def\sbr {\tilde{b}_R}
\def\msbl{m_{\sbl}}
\def\msbr{m_{\sbr}}
\def\sba {\tilde{b}_1}
\def\sbb {\tilde{b}_2}
\def\msba{m_{\sba}}
\def\msbb{m_{\sbb}}
\def\thsb{\theta_{\tilde{b}}}

\newcommand{\sfa}{\tilde{f}}

\newcommand{\SLASH}[2]{\makebox[#2ex][l]{$#1$}/}
\newcommand{\kslash}{\SLASH{k}{.15}}
\newcommand{\pslash}{\SLASH{p}{.2}}
\newcommand{\qslash}{\SLASH{q}{.08}}

\def\d  {{\rm d}}
\def\eps{\varepsilon}
\def\O  {{\cal O}}

\def\beq{\begin{equation}}
\def\eeq{\end{equation}}
\def\bea{\begin{eqnarray}}
\def\eea{\end{eqnarray}}

\preprint{DESY 02-207}
\preprint{hep-ph/0303032}
\title{Gluino Pair Production in High-Energy Photon Collisions}
\author{Stefan Berge}
\affiliation{{II.} Institut f\"ur Theoretische Physik, Universit\"at Hamburg,
             Luruper Chaussee 149, D-22761 Hamburg, Germany}
\author{Michael Klasen}
\email[]{michael.klasen@desy.de}
\affiliation{{II.} Institut f\"ur Theoretische Physik, Universit\"at Hamburg,
             Luruper Chaussee 149, D-22761 Hamburg, Germany}
\date{\today}
\begin{abstract}
We study the potential of high-energy photon colliders for the production of
gluino pairs within the Minimal Supersymmetric Standard Model (MSSM). In this
model, the process $\gamma\gamma\to\go\go$ is mediated by quark/squark box
diagrams with enhancements for up-type quarks/squarks from their larger charges
and for third generation squarks from their large mass splittings, generated by
the mixing of left- and right-handed states. Far above threshold and in
scenarios with very heavy squarks, resolved photons can contribute
significantly at tree level. Taking into account the laser photon
backscattering spectrum, electron and laser beam polarization effects, and
current mass exclusion limits, we find that gluino pair production in
high-energy photon collisions should be visible over large regions of the MSSM
parameter space, contrary to what has been found for $e^+e^-$ annihilation. In
addition, the cross section rises rather steeply, so that a gluino mass
determination with a precision of a few GeV should be feasible for a wide
range of post-LEP benchmark points.
\end{abstract}
\maketitle


\section{Introduction}
\label{sec:1}

Weak-scale supersymmetry (SUSY) is one of the most attractive extensions of the
Standard Model (SM) of particle physics \cite{Nilles:1983ge,Haber:1984rc}. If
it is realized in nature, SUSY particles will be discovered either at Run II
of the Fermilab Tevatron \cite{Carena:2000yx,Abel:2000vs,Culbertson:2000am,
Ambrosanio:1998zf,Allanach:1999bf} or within the first years of running at the
CERN LHC \cite{:1999fr,Abdullin:1998pm}. Reconstruction of the SUSY Lagrangian
and a precise determination of its free parameters will, however, require the
clean environment of a linear $e^+e^-$ collider, where in particular the
masses, phases, and (electroweak) couplings of sfermions and gauginos will be
determined with high accuracy \cite{Aguilar-Saavedra:2001rg}. However, the
mass and coupling of the gluino will pose some difficulties, since the gluino
couples only to strongly interacting particles and is thus produced only at
the one-loop level or in multi-parton final states.

In a recent publication, we have investigated gluino pair production through
triangular quark/squark loops in $e^+e^-$ annihilation with center-of-mass
energies up to 3 TeV, which may become available in the future at linear
colliders like DESY TESLA or CERN CLIC \cite{Berge:2002ev,Berge:2002xc}. Due
to large cancellation effects, we found that promisingly large cross sections
can only be expected for scenarios with large left-/right-handed up-type
squark mass splittings or with large top-squark mixing and for gluino masses
up to 500 GeV. Small gluino masses of 200 GeV might be measured with a
precision of about 5 GeV in center-of-mass energy scans with luminosities of
100 fb$^{-1}$/point. However, when both the left-/right-handed squark mass
splitting and the squark mixing remain small, gluino pair production in
$e^+e^-$ annihilation will be hard to observe, even with luminosities of 1000
fb$^{-1}$/year.

In this Paper, we study the potential of high-energy photon colliders for the
production of gluino pairs within the Minimal Supersymmetric Standard Model
(MSSM). Similar studies have previously been carried out for the production of
SM \cite{Jikia:1999en,Melles:1999xd,Niezurawski:2002aq}, MSSM \cite{
Muhlleitner:2001kw}, and double-charged Higgs bosons \cite{Chakrabarti:1998qy},
sfermion pairs \cite{Berge:2000cb,Klasen:2000cc}, and various other processes
\cite{Badelek:2001xb}. In the MSSM,
the process $\gamma\gamma\to\go\go$ is mediated by quark/squark box
diagrams with enhancements for up-type quarks/squarks from their larger charges
and for third generation squarks from their large mass splittings, generated by
the mixing of left- and right-handed states. At tree level, gluinos can be
produced in pairs only in association with two quarks, or they are produced
singly in association with a quark and a squark. Both processes result in
multi-jet final states, where phase space is limited and gluinos may be hard
to isolate. Far above threshold and in scenarios with very heavy squarks,
resolved photons can contribute significantly at tree level. Taking into
account the usual laser photon backscattering spectrum \cite{Ginzburg:1982yr},
electron and laser beam polarization effects, and current mass exclusion
limits, we find that gluino pair production in high-energy photon collisions
should be visible over large regions of the MSSM parameter space, contrary to
what has been found for $e^+e^-$ annihilation. In addition, the cross section
rises rather steeply, so that a gluino mass determination with a precision of
a few GeV should be feasible for a wide range of post-LEP benchmark points.

Our calculations involve various masses and couplings of SM particles,
for which we use the most up-to-date values from the 2002 Review of
the Particle Data Group \cite{Hagiwara:pw}. In particular, we evaluate the
electromagnetic fine structure constant $\alpha(m_Z)=1/127.934$ at the mass of
the $Z^0$-boson, $m_Z = 91.1876$ GeV, and calculate the weak mixing angle
$\theta_W$ from the tree-level expression $\sin^2\theta_W=1-m_W^2/m_Z^2$ with
$m_W = 80.423 $ GeV. Among the quark masses, only the one of the top quark,
$m_t = 174.3$ GeV, plays a significant role, while the bottom quark mass, $m_b
= 4.7$ GeV, and the charm quark mass, $m_c = 1.5$ GeV, could have been
neglected like those of the three light quarks. The strong coupling constant is
evaluated at the gluino mass scale from the one-loop expression with five
active flavors and $\Lambda_{\rm LO}^{n_f=5} = 83.76$ MeV, corresponding to
the world avarage (available only at two loops) of $\alpha_s(m_Z) = 0.1172$. A
variation of the renormalization scale by a factor of four about the gluino
mass results in a cross section uncertainty of about $\pm25$ \%, which can be
reduced considerably by including full next-to-leading order QCD corrections.
Like the heavy top quark, all SUSY particles have been decoupled from the
running of the strong coupling constant.

We work in the framework of the MSSM with conserved $R$- (matter-) parity,
which represents the simplest phenomenologically viable model, but which is
still sufficiently general to not depend on a specific SUSY breaking mechanism.
Models with broken $R$-parity are severely restricted by the non-observation
of proton decay, which would violate both baryon and lepton number
conservation. We do not consider light gluino mass windows, which may or may
not be excluded from searches at fixed target and collider experiments
\cite{Hagiwara:pw}. Instead, we adopt the current mass limit $\mg\geq 200$ GeV
from the CDF \cite{Affolder:2001tc} and D0 \cite{Abachi:1995ng} searches in the
jets with missing energy channel, relevant for non-mixing squark masses of
$\ms\geq 325$ GeV and $\tan\beta = 3$. Values for the ratio of the Higgs vacuum
expectation values, $\tan\beta$, below 2.4 are already excluded by the CERN LEP
experiments, although this value is obtained using one-loop corrections only
and depends in addition on the top quark mass. Furthermore, values of
$\tan\beta$ between 2.4 and 8.5 are only allowed in a very narrow window of
light Higgs boson masses between 113 and 127 GeV \cite{lhwg:2001xx}. Therefore,
we employ a safely high value of $\tan\beta = 10$.
If not stated otherwise, we adopt the smallest allowed universal squark mass of
$\ms\simeq\msusy=325$ GeV and large top-squark mixing with
$\thst=45.195^\circ$, $\msta=110.519$ GeV, and $\mstb=505.689$ GeV,
which can be generated by choosing appropriate values for the Higgs mass
parameter, $\mu=-500$ GeV, and the trilinear top-squark coupling, $A_t=648.512$
GeV \cite{Hahn:2001rv}. The SUSY one-loop contributions to the $\rho$-parameter
and the light top-squark mass $\msta$ are then still significantly below and
above the CERN LEP limits, $\rho_{\rm SUSY} < 0.0012^{+0.0023}_{-0.0014}$ and
$\msta\geq 100$ GeV \cite{Hagiwara:pw,lswg:2002xx}. For small and intermediate
values of $\tan\beta$, mixing in the bottom squark sector remains small, and
we take $\thsb=0^\circ$ as for the four light squark flavors.

The remainder of this Paper is organized as follows: In Sec.\ \ref{sec:2} we
compute the one-loop amplitudes for gluino pair production in direct
photon-photon collisions, discuss briefly their analytical properties, and
study in detail their numerical dependence on the squark masses and mixing
angles, on the photon polarization and center-of-mass energy. In Sec.\
\ref{sec:3} we calculate the squared matrix elements for the resolved photon
contributions from quark-antiquark and gluon-gluon scattering analytically
and discuss several typical cases, where resolved processes can be numerically
important. In Sec. \ref{sec:4} we present expected total cross sections for
gluino pair production with laser-backscattered photons in polarized
electron-electron collisions for various post-LEP benchmark SUSY scenarios.
We then estimate the precision with which the gluino mass might be determined
for various typical squark and gluino masses and realistic photon collider
luminosities. Finally, our conclusions are given in Sec.\ \ref{sec:5}.

\section{Direct Photon-Photon Scattering}
\label{sec:2}

\begin{figure}
 \centering
 \epsfig{file=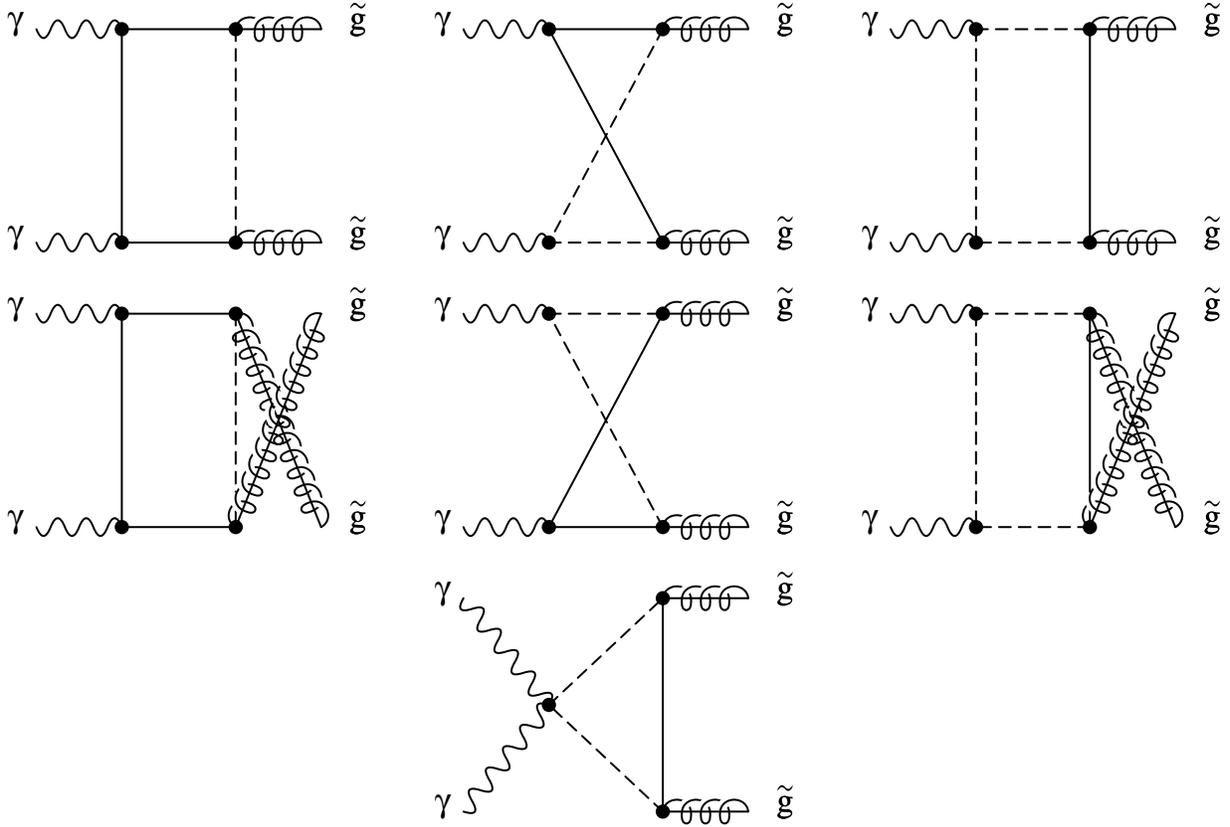,bbllx=60pt,bblly=217pt,bburx=374pt,bbury=433pt,%
  width=\columnwidth}
 \caption{\label{fig:7}Feynman diagrams for gluino pair production in direct
 photon-photon collisions. The photons couple to the produced gluinos through
 quark (full lines) and squark (dashed lines) box diagrams with flavor flow in
 both directions.}
\end{figure}

As the supersymmetric partners of the gauge bosons of the strong interaction,
gluinos couple only to colored particles and sparticles and can therefore not
be pair-produced directly at tree-level in photon-photon collisions. Instead,
they are produed at the one-loop level through the box diagrams shown in Fig.\
\ref{fig:7}, where the virtual quark/squark flavor can flow in both
%
%
directions. Topologically, additional triangular loop diagrams without a
quartic vertex and two-point functions with a quartic vertex are also allowed,
but they evaluate to zero, since they involve internal gluon propagators and
only one (traceless) color matrix inside the quark/squark loop. A full set
of diagrams can be generated with the computer algebra package FeynArts
\cite{Kublbeck:xc,Hahn:2000kx}.

Denoting the four-momenta of the incoming photons with $p_i^\mu$, their
polarisation vectors with $\varepsilon^\mu(p_i)$, the four-momenta of the
produced gluinos with $k_i^\mu$, and employing the Feynman rules of SUSY-QCD
\cite{Haber:1984rc}, we can write down the scattering amplitudes for the
diagrams with one internal squark propagator,
\begin{eqnarray}
{\cal M}_{t1}&=& \int {\d^4q\over(2\pi)^4} i \varepsilon^{\mu}(p_1)\varepsilon^{\nu}(p_2) \Big[\overline{u}(k_1,m_{\tilde{g}}) \left(-i {\sqrt{2}}g_{s}\omega_{-}T_{mn}^{a}S_{i1}^{\tilde{q}} + i {\sqrt{2}}g_{s}\omega_{+}T_{mn}^{a}S_{i2}^{\tilde{q}}\right)\nonumber\\
&\times& \left( -\pslash_2 - \qslash + \kslash_1 + \kslash_2+m_{q}\right)\left(-iee_q\gamma^{\mu}\right)\left(-\qslash + m_{q}\right) \left(-iee_q\gamma^{\nu}\right)\left(-\pslash_2 - \qslash + m_{q}\right)\nonumber\\
&\times& \left(i {\sqrt{2}}g_{s}{S_{i2}^{\tilde{q}}}^{\dagger}\omega_{-}T_{nm}^{b} -i {\sqrt{2}}g_{s}{S_{i1}^{\tilde{q}}}^{\dagger}\omega_{+}T_{nm}^{b}\right) v(k_2,m_{\tilde{g}}) \Big] / \Big[ \left({q}^2 - {m_{q}}^2\right) \left({\left( p_2 + q \right) }^2 - {m_{q}}^2\right)\nonumber\\
&\times& \left({\left( p_2 + q - k_2 \right) }^2 - {m_{\tilde{q}_{i}}}^2\right)\left({\left( p_2 + q - k_1 - k_2 \right) }^2 - {m_{q}}^2\right)\Big],\label{eq:1}\\
&&\nonumber\\
{\cal M}_{u1}&=& \int {\d^4q\over(2\pi)^4} i \varepsilon^{\mu}(p_1)\varepsilon^{\nu}(p_2) \Big[\overline{u}(k_1,m_{\tilde{g}}) \left(i {\sqrt{2}}g_{s}{S_{i2}^{\tilde{q}}}^{\dagger}\omega_{-}T_{mn}^{a} - i {\sqrt{2}}g_{s}{S_{i1}^{\tilde{q}}}^{\dagger}\omega_{+}T_{mn}^{a}\right)\nonumber\\
&\times&\left(\pslash_2 + \qslash+ m_{q}\right)\left(iee_q\gamma^{\nu}\right)\left(\qslash + m_{q}\right)\left(iee_q\gamma^{\mu}\right)\left(\pslash_2 + \qslash - \kslash_1 - \kslash_2 + m_{q}\right)\nonumber\\
&\times&\left(-i {\sqrt{2}}g_{s}\omega_{-}T_{nm}^{b}S_{i1}^{\tilde{q}}+ i {\sqrt{2}}g_{s}\omega_{+}T_{nm}^{b}S_{i2}^{\tilde{q}}\right)v(k_2,m_{\tilde{g}})\Big]/\Big[\left({q}^2 - {m_{q}}^2\right)\left({\left( p_2 + q \right) }^2 - {m_{q}}^2\right) \nonumber\\
&\times&\left({\left( p_2 + q - k_1 \right) }^2 - {m_{\tilde{q}_{i}}}^2\right)\left({\left( p_2 + q - k_1 - k_2 \right) }^2 - {m_{q}}^2\right)\Big],
\end{eqnarray}
for the diagrams with two internal squark propagators,
\begin{eqnarray}
{\cal M}_{t2}&=& \int {\d^4q\over(2\pi)^4} e e_q (-p_2 - 2q + 2k_1)^{\nu}\varepsilon^{\mu}(p_1)\varepsilon^{\nu}(p_2) \Big[\overline{u}(k_1,m_{\tilde{g}})\nonumber\\
&\times&\left(i {\sqrt{2}}g_{s}{S_{i2}^{\tilde{q}}}^{\dagger}\omega_{-}T_{mn}^{a} - i {\sqrt{2}}g_{s}{S_{i1}^{\tilde{q}}}^{\dagger}\omega_{+}T_{mn}^{a}\right) \left(\qslash + m_{q}\right)\left(iee_q\gamma^{\mu} \right)\left(\pslash_2 + \qslash - \kslash_1 - \kslash_2 + m_{q}\right)\nonumber\\
&\times& \left(-i {\sqrt{2}}g_{s}\omega_{-}T_{nm}^{b}S_{i1}^{\tilde{q}} + i {\sqrt{2}}g_{s}\omega_{+}T_{nm}^{b}S_{i2}^{\tilde{q}}\right)v(k_2,m_{\tilde{g}})\Big]/\Big[\left({q}^2 - {m_{q}}^2\right)\left({\left( q - k_1 \right) }^2 - {m_{\tilde{q}_{i}}}^2\right)\nonumber\\
&\times& \left({\left( p_2 + q - k_1 \right) }^2 - {m_{\tilde{q}_{i}}}^2\right)\left({\left( p_2 + q - k_1 - k_2 \right) }^2 - {m_{q}}^2\right)\Big], \\
&&\nonumber\\
{\cal M}_{u2}&=& \int {\d^4q\over(2\pi)^4} (-e e_q) (p_2 + 2q - k_1 - k_2)^{\mu}\varepsilon^{\mu}(p_1)\varepsilon^{\nu}(p_2)\Big[\overline{u}(k_1,m_{\tilde{g}}) \nonumber\\
&\times& \left(-i {\sqrt{2}}g_{s}\omega_{-}T_{mn}^{a}S_{i1}^{\tilde{q}} + i {\sqrt{2}}g_{s}\omega_{+}T_{mn}^{a}S_{i2}^{\tilde{q}}\right)\left(-\qslash + \kslash_1 + m_{q}\right)\left(-iee_q\gamma^{\nu} \right)\left(-\pslash_2 - \qslash + \kslash_1 + m_{q}\right)\nonumber\\
&\times&  \left(i {\sqrt{2}}g_{s}{S_{i2}^{\tilde{q}}}^{\dagger}\omega_{-}T_{nm}^{b} - i {\sqrt{2}}g_{s}{S_{i1}^{\tilde{q}}}^{\dagger}\omega_{+}T_{nm}^{b}\right)v(k_2,m_{\tilde{g}})\Big]/\Big[\left({q}^2 - {m_{\tilde{q}_{i}}}^2\right)\left( {\left( q - k_1 \right) }^2 - {m_{q}}^2\right)\nonumber\\
&\times& \left({\left( p_2 + q - k_1 \right) }^2 - {m_{q}}^2\right)\left({\left( p_2 + q - k_1 - k_2 \right) }^2 - {m_{\tilde{q}_{i}}}^2\right)\Big],\\
&&\nonumber\\
{\cal M}_{x2}&=& \int {\d^4q\over(2\pi)^4} 2 i e^2 e_q^2 g^{\mu\nu}\varepsilon^{\mu}(p_1)\varepsilon^{\nu}(p_2) \Big[\overline{u}(k_1,m_{\tilde{g}})\left(-i {\sqrt{2}}g_{s}\omega_{-}T_{mn}^{a}S_{i1}^{\tilde{q}} + i {\sqrt{2}}g_{s}\omega_{+}T_{mn}^{a}S_{i2}^{\tilde{q}}\right)\nonumber\\
&\times& \left(-\qslash + m_{q}\right)\left(i {\sqrt{2}}g_{s}{S_{i2}^{\tilde{q}}}^{\dagger}\omega_{-}T_{nm}^{b} - i {\sqrt{2}}g_{s}{S_{i1}^{\tilde{q}}}^{\dagger}\omega_{+}T_{nm}^{b}\right)v(k_2,m_{\tilde{g}})\Big]/\Big[\left({q}^2 - {m_{q}}^2\right)\nonumber\\
&\times& \left({\left( q + k_1 \right) }^2 - {m_{\tilde{q}_{i}}}^2\right)\left({\left( q - k_2 \right) }^2 - {m_{\tilde{q}_{i}}}^2\right)\Big],
\end{eqnarray}
and for the diagrams with three internal squark propagators,
\begin{eqnarray}
{\cal M}_{t3}&=& \int {\d^4q\over(2\pi)^4} (-i e^2 e_q^2)(-p_2 - 2q)^{\nu}(-p_2 - 2q + k_1 + k_2)^{\mu}\varepsilon^{\mu}(p_1)\varepsilon^{\nu}(p_2) \Big[\overline{u}(k_1,m_{\tilde{g}})\nonumber\\
&\times& \left(i {\sqrt{2}}g_{s}{S_{i2}^{\tilde{q}}}^{\dagger}\omega_{-}T_{mn}^{a}+-i {\sqrt{2}}g_{s}{S_{i1}^{\tilde{q}}}^{\dagger}\omega_{+}T_{mn}^{a}\right)\left(\pslash_2 + \qslash - \kslash_2 + m_{q}\right)\nonumber\\
&\times& \left(-i {\sqrt{2}}g_{s}\omega_{-}T_{nm}^{b}S_{i1}^{\tilde{q}}+i {\sqrt{2}}g_{s}\omega_{+}T_{nm}^{b} S_{i2}^{\tilde{q}}\right)v(k_2,m_{\tilde{g}})\Big]/\Big[\left({q}^2 - {m_{\tilde{q}_{i}}}^2\right)\left({\left( p_2 + q \right) }^2 - {m_{\tilde{q}_{i}}}^2\right) \nonumber\\
&\times& \left({\left( p_2 + q - k_2 \right) }^2 - {m_{q}}^2\right)\left({\left( p_2 + q - k_1 - k_2 \right) }^2 - {m_{\tilde{q}_{i}}}^2\right),\ {\rm and}\\
&&\nonumber\\
{\cal M}_{u3}&=& \int {\d^4q\over(2\pi)^4} (-i e^2 e_q^2)(-p_2 - 2q)^{\nu}(-p_2 - 2q + k_1 + k_2)^{\mu}\varepsilon^{\mu}(p_1)\varepsilon^{\nu}(p_2) \Big[\overline{u}(k_1,m_{\tilde{g}})\nonumber\\
&\times& \left(-i {\sqrt{2}}g_{s}\omega_{-}T_{mn}^{a}S_{i1}^{\tilde{q}}+i {\sqrt{2}}g_{s}\omega_{+}T_{mn}^{a} S_{i2}^{\tilde{q}}\right)\left(-\pslash_2 - \qslash + \kslash_1 + m_{q}\right)\nonumber\\
&\times&\left(i {\sqrt{2}}g_{s}{S_{i2}^{\tilde{q}}}^{\dagger}\omega_{-}T_{nm}^{b}+-i {\sqrt{2}} g_{s}{S_{i1}^{\tilde{q}}}^{\dagger}\omega_{+}T_{nm}^{b}\right)v(k_2,m_{\tilde{g}})\Big]/\Big[\left({q}^2 - {m_{\tilde{q}_{i}}}^2\right)\left({\left( p_2 + q \right) }^2 - {m_{\tilde{q}_{i}}}^2\right) \nonumber\\
&\times& \left({\left( p_2 + q - k_1 \right) }^2 - {m_{q}}^2\right)\left({\left( p_2 + q - k_1 - k_2 \right) }^2 - {m_{\tilde{q}_{i}}}^2\right)\Big]. \label{eq:7}
\end{eqnarray}
Here, $e$ and $g_s$ are the electromagnetic and strong couplings, respectively,
$e_q$ is the fractional charge of the quark flavor $q$ in the loop, and
$T_{mn}^a$ is the SU(3) color matrix attached to gluinos of color $a$, and
summation over the (s)quark color indices $m,n$ is implied. $\mg$, $m_q$, and
$m_{\tilde{q}_i}$ are the masses of the gluino, quark, and squark mass
eigenstate $i$, and $q^\mu$ is the internal loop momentum. For diagrams with
opposite flavor flow, the projectors $\omega_{\pm}=(1\pm \gamma_5)/2$ must be
interchanged and the Hermitean conjugate of the squark mixing matrices
$S^{\tilde{q}}_{ij}$ must be taken. The loop integrals in Eqs.\
(\ref{eq:1})-(\ref{eq:7}) are free of ultraviolet divergences, since they
involve at least three propagators and no tree-level coupling of the photon to
the gluino that would require renormalization. They are also free of infrared
and collinear singularities, since massless gluons do not appear, and are
therefore most easily evaluated numerically \cite{vanOldenborgh:1989wn,
vanOldenborgh:1990yc,Hahn:1998yk}.

After summing these amplitudes over internal quark flavors $q$, squark mass
eigenstates $i$, colors and helicities of the produced gluinos, we obtain the
direct photon-photon cross section
\beq
 \sigma^{\rm dir}_{\gamma\gamma} =
 {1\over2s_{\gamma\gamma}}{1\over8\pi s_{\gamma\gamma}}
 {1\over2} \int \d\tg \sum_{a,b}|{\cal M}_{\gamma\gamma}|^2, \label{eq:8}
\eeq
where 
\bea
 {\cal M}_{\gamma\gamma} &=& \sum_{q=1}^{6} \sum_{i=1}^{2} \left(
 {\cal M}_{t1}+{\cal M}_{u1}+{\cal M}_{t2}+{\cal M}_{u2}+{\cal M}_{x2}+
 {\cal M}_{t3}+{\cal M}_{u3} \right) \nonumber\\
 &+&(\omega_+\leftrightarrow\omega_-\,,\,{S_{ij}^{\tilde{q}}}\leftrightarrow
 {S_{ij}^{\tilde{q}}}^{\dagger}),
\eea
$s_{\gamma\gamma}=(p_1+p_2)^2$, $t_{\go}=(p_1-k_1)^2-\mg^2$, and $u_{\go}=
(p_1-k_2)-\mg^2$ are the mass-subtracted Lorentz-invariant Mandelstam
variables, $t_{\go}$ is integrated in the range $[-s_{\gamma\gamma}\pm
\sqrt{s_{\gamma\gamma}(s_{\gamma\gamma}-4\mg^2)}]/2$,
and where we have included a factor of 1/2 for the production of
two identical Majorana fermions. For unpolarized cross sections, we sum in
addition over the transverse polarizations $T$ of the initial photons with the
completeness relation
\beq
 \sum_{T}\varepsilon^{\mu\ast}(p_i)\varepsilon^{\nu}(p_i)=-g^{\mu\nu}
\eeq
and include a spin averaging factor of 1/2 for each photon.

\begin{figure}
 \centering
 \epsfig{file=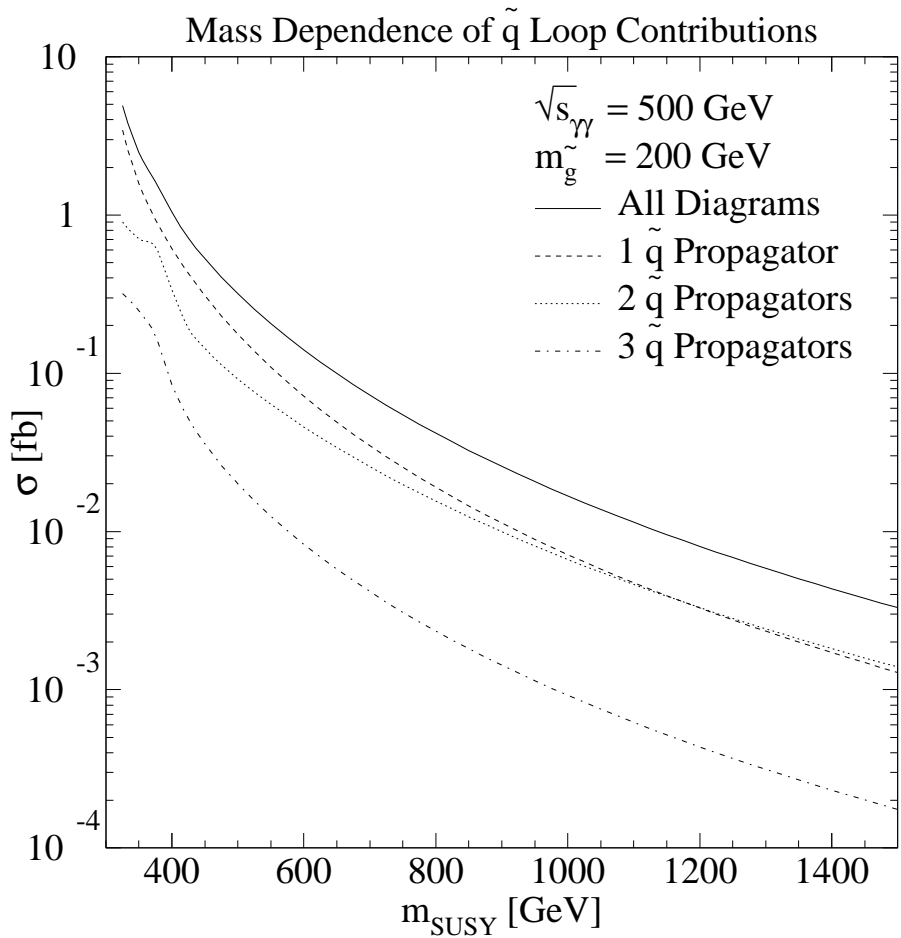,width=0.8\columnwidth}
 \caption{\label{fig:5}Dependence of the gluino pair production cross section
 in unpolarized photon-photon collisions on the squark mass parameter $\msusy$.
 The numerical importance of the amplitudes decreases with the number of
 internal squark propagators.}
\end{figure}

As is evident from Fig.\ \ref{fig:5}, the numerical importance of the
%
%
amplitudes in Eqs.\ (\ref{eq:1})-(\ref{eq:7}) depends directly on the number
of internal squark propagators, which induce a suppression with the heavy
squark mass $\ms$ or, equivalently, the parameter $\msusy$. The dominant
contribution comes from diagrams with one internal squark propagator, but
diagrams with two internal squark propagators still contribute significantly
and can not be neglected as proposed in Ref.\ \cite{Li:hu}. They are even more
important once the photon-photon center-of-mass energy allows for the
production of two intermediate on-shell squarks, {\it i.e.} when
$\sqrt{s}_{\gamma\gamma}>2\ms$, so that real and imaginary parts of the loop
diagrams contribute. Only the diagrams with three internal squark propagators
are indeed of negligible impact.

\begin{figure}
 \centering
 \epsfig{file=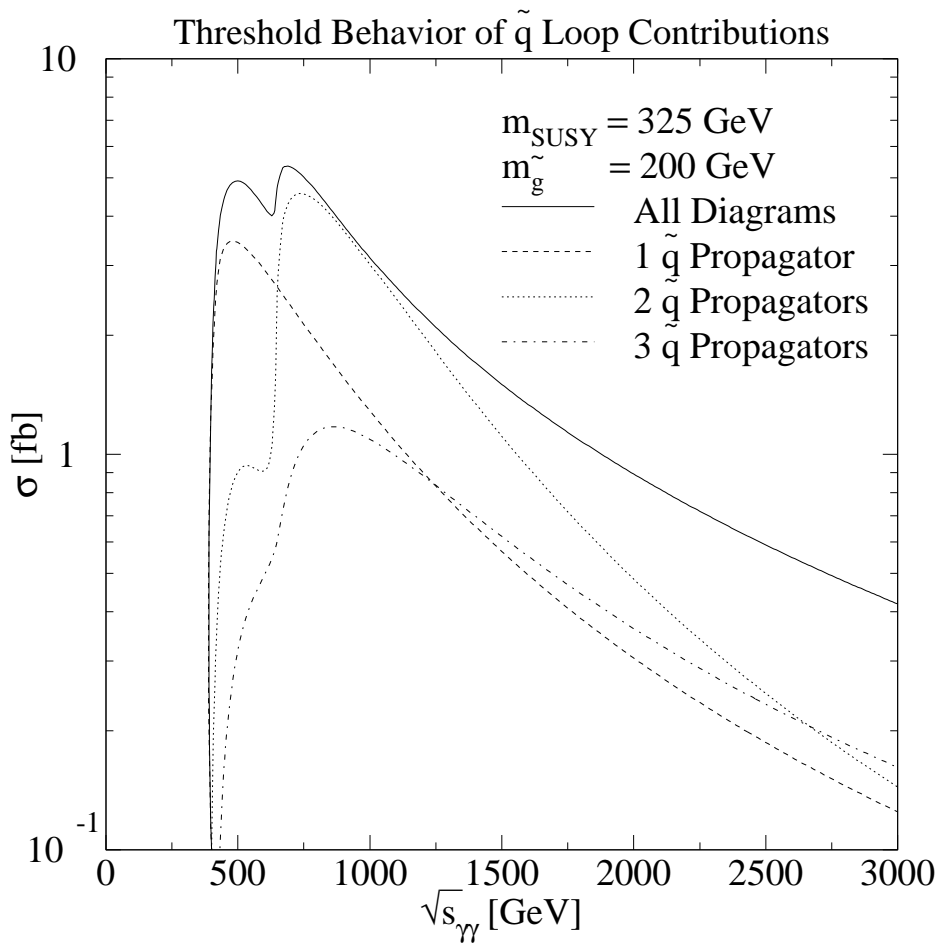,width=0.8\columnwidth}
 \caption{\label{fig:12}Threshold behavior of the gluino pair production cross
 section in unpolarized photon-photon collisions for diagrams involving
 different numbers of squark propagators.}
\end{figure}

The number and ($s$- or $t$-channel) nature of the squark propagators occuring
in the Feynman diagrams of Fig.\ \ref{fig:7} is clearly reflected in a distinct
threshold behavior, shown in Fig.\ \ref{fig:12}. For example, the diagrams in
%
%
the center column of Fig.\ \ref{fig:7} involve two $s$-channel squark
propagators, which can become on-shell when $\sqrt{s}_{\gamma\gamma}\geq 2
\ms\simeq2\msusy$. The same observation holds for the diagrams in the right
column of Fig.\ \ref{fig:7}, which are, however, suppressed by an additional
$t$-channel squark propagator. For lighter gluinos with mass $\mg < m_t$ or
$\mg < \msta$, similar resonance structures would be visible where the
center-of-mass energy crosses the pair production threshold for top quarks and
squarks.
For very light gluinos of mass $\mg=5...25$ GeV, squarks of mass $50...150$
GeV, and neglecting the diagrams with more than one squark propagator, all
quark masses, and squark mixing effects, we can roughly reproduce the shapes
and magnitudes (in pb, not nb) of the threshold behavior in Fig.\ 2 of Ref.\
\cite{Li:hu}. For a more detailed numerical comparison, more information on the
quark charges and coupling constants used there would be needed.

\begin{figure}
 \centering
 \epsfig{file=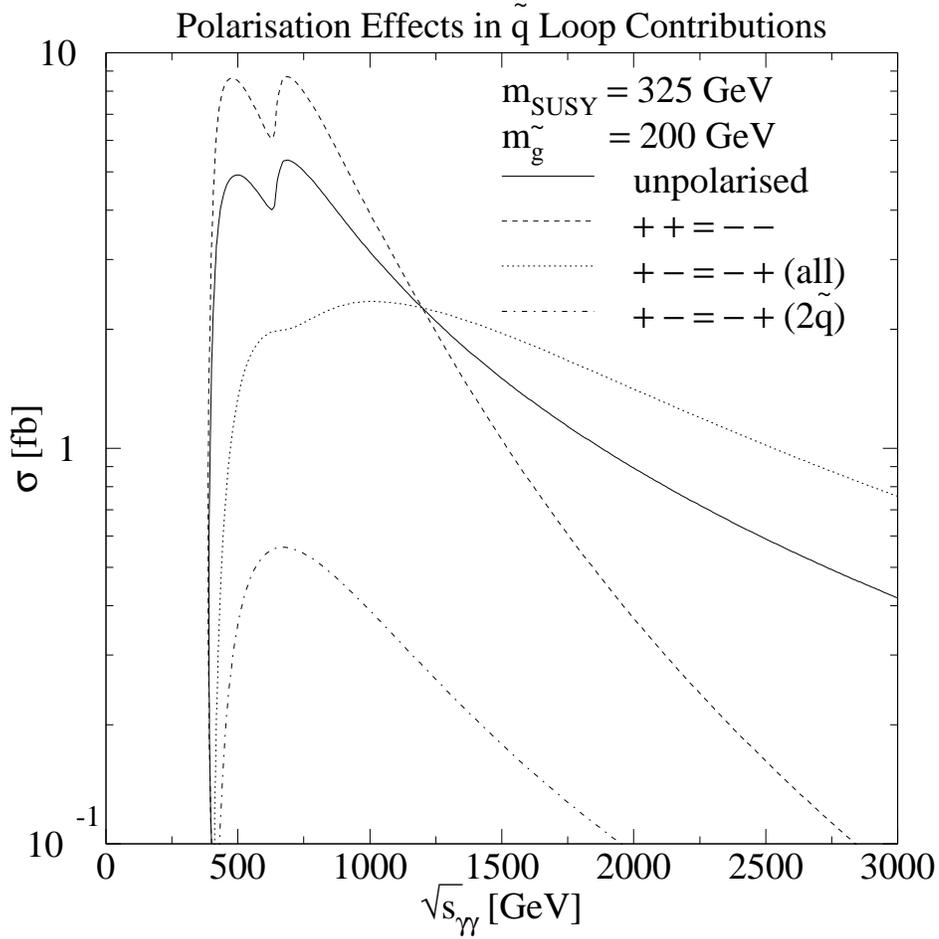,width=0.8\columnwidth}
 \caption{\label{fig:11}Threshold behavior of the gluino pair production cross
 section in polarized photon-photon collisions.}
\end{figure}

As can be seen in Fig.\ \ref{fig:11}, the threshold behavior of the gluino
%
%
pair production cross section is also influenced by the polarisation of the
initial photons. If they carry the same helicity, the cross section shows the
steep rise typical for $S$-wave production of the two gluinos, whereas for
photons with opposite helicity the gluinos are produced as a $P$-wave, the
cross section rises more slowly, and the intermediate squarks in diagrams with
two $s$-channel propagators can not be produced on-shell (see dot-dashed curve
in Fig.\ \ref{fig:11}). For center-of-mass energy scans it is therefore
advantageous to choose the same helicities for both photons.

\begin{figure}
 \centering
 \epsfig{file=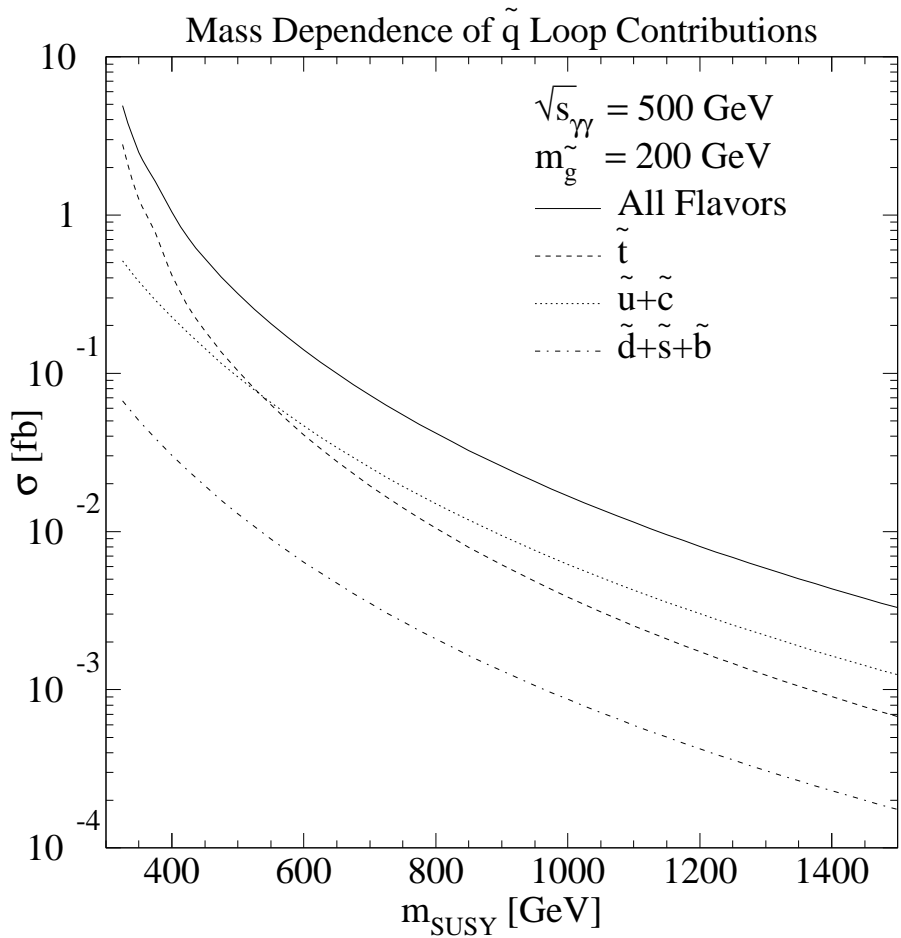,width=0.8\columnwidth}
 \caption{\label{fig:6}Dependence of the gluino pair production cross section
 in unpolarized photon-photon collisions on the squark mass parameter $\msusy$.
 Down-type ($d$, $s$, and $b$) quarks and squarks are suppressed relative to
 up-type ($u$, $c$, and $t$) quarks and squarks due to their smaller
 electromagnetic charges.}
\end{figure}

Since the cross section in Eq.\ (\ref{eq:8}) is proportional to the fourth
power of the fractional quark charge, up-type (s)quarks contribute 16 times as
much as down-type (s)quarks (see Fig.\ \ref{fig:6}). In addition, the
%
%
top-squark contribution is enhanced for small $\msusy$ due to the large
top-squark mixing angle and the small mass of the light top-squark mass
eigenstate.

\begin{figure}
 \centering
 \epsfig{file=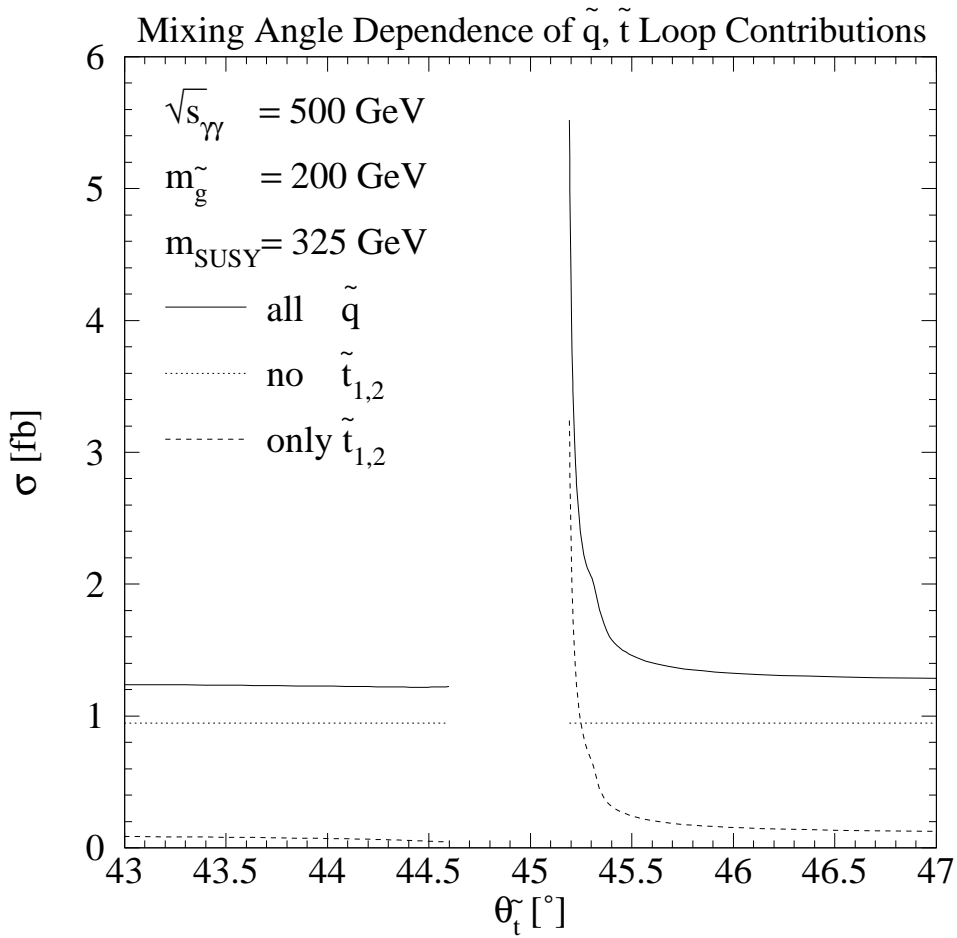,width=0.8\columnwidth}
 \caption{\label{fig:10}Dependence of the gluino pair production cross section
 in unpolarized photon-photon collisions on the top-squark mixing angle.}
\end{figure}

The influence of the top-squark mixing angle is shown in Fig.\ \ref{fig:10},
%
%
where the central region is excluded by the CERN LEP limits on the
$\rho$-parameter and $\msta$ \cite{Hagiwara:pw,lswg:2002xx}. It is quite strong
above the excluded region, $\thst\geq45.192^\circ$, where the mass of the light
top-squark mass eigenstate rises quickly from $\msta\geq 100$ GeV to values
around $\msusy=325$ GeV. In this region, the top-squark contribution (dashed
curve) to the gluino pair production cross
section drops by more than one order of magnitude, whereas the total cross
section (full curve) drops only by about a factor of four due to the
interference with the constant contributions from the other, non-mixing
(s)quarks (dotted curve).

\section{Resolved Photon-Photon Scattering}
\label{sec:3}

As the gauge boson of the electromagnetic interaction, the photon can only
interact directly with charged (s)particles, so that gluino pair production
arises only at one-loop and $\O(\alpha^2\alpha_s^2)$. However, in high-energy
scattering processes the photon can also fluctuate into intermediate
quark-antiquark states and develop a complex hadronic structure. The quarks
and gluons inside the photon carry then only a fraction of the photon's energy.
When both photons are resolved into quarks or gluons, the latter can interact
strongly and gluino pairs can be produced already at tree-level. These
$\O(\alpha_s^2)$ double-resolved processes can be numerically large, despite
the fact that a suppression from the $\O(\alpha/\alpha_s)$ quark and
$\O(\alpha)$ gluon densities in the photon must be taken into account. If only
one photon is resolved, gluino pairs are produced through the one-loop diagrams
in Fig.\ \ref{fig:7} with one photon being replaced by a gluon (and the
additional color-octet diagrams not shown in Fig.\ \ref{fig:7}), so that
they are then of $\O(\alpha\alpha_s^3)$. Due to their additional suppression
from the $\O(\alpha)$ gluon density, these contributions are numerically small
and will be neglected in the following.

\begin{figure}
 \centering
 \epsfig{file=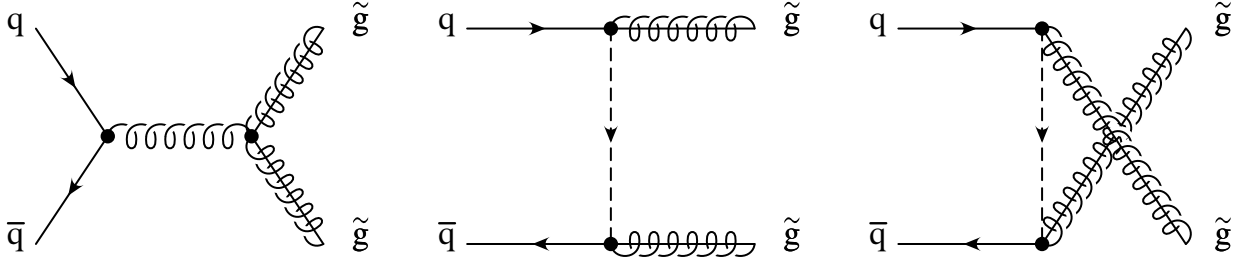,bbllx=62pt,bblly=360pt,bburx=373pt,bbury=432pt,%
  width=\columnwidth}
 \caption{\label{fig:9}Feynman diagrams for gluino pair production in
 quark-antiquark collisions.}
\end{figure}

\begin{figure}
 \centering
 \epsfig{file=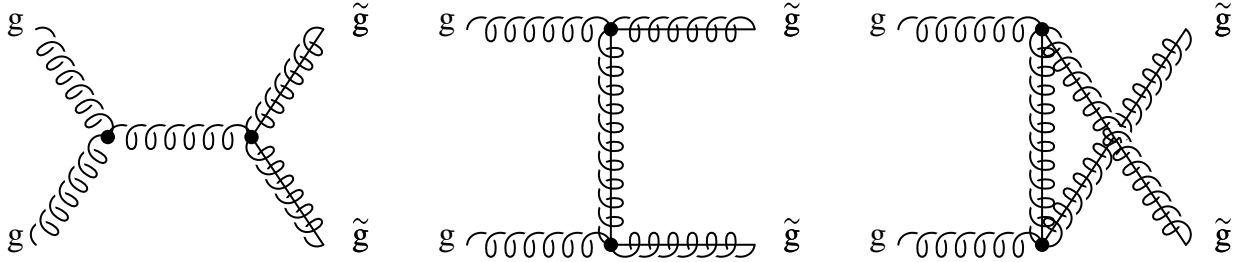,bbllx=62pt,bblly=360pt,bburx=373pt,bbury=432pt,%
  width=\columnwidth}
 \caption{\label{fig:8}Feynman diagrams for gluino pair production in
 gluon-gluon collisions.}
\end{figure}

Gluino pair production in double-resolved photon-photon scattering arises from
the quark-antiquark and gluon-gluon scattering diagrams shown in Figs.\
\ref{fig:9} and \ref{fig:8}.
%
%
%
%
The corresponding spin- and color-averaged squared matrix elements are
\cite{Dawson:1983fw}
\bea
 \overline{|{\cal M}_{q\bar{q}}|}^2 &=&
 {2g_s^4 (N_C^2-1)\ N_C\over4N_C^2} \left( 2{2\mg^2 s+\tg^2+\ug^2\over s^2}
 +2{\mg^2s+\tg^2\over s\ts}+2{\mg^2s+\ug^2\over s\us}
 +{\tg^2\over\ts^2}+{\ug^2\over\us^2}\right)\quad \label{eq:11}\\
 &+&{2g_s^4 (N_C^2-1)/N_C\over 4N_C^2} \left( 2{\mg^2 s\over\ts\us}
 -{\tg^2\over\ts^2}-{\ug^2\over\us^2}\right)\ {\rm and}\nonumber\\
&&\nonumber\\
 \overline{|{\cal M}_{gg}|}^2 &=&
 {8g_s^4 (N_C^2-1)\ N_C^2\over4(N_C^2-1)^2}\left(1-{\tg\ug\over s^2}\right)
 \left[{s^2\over\tg\ug}-2+4{\mg^2s\over\tg\ug}\left(1-{\mg^2s\over\tg\ug}
 \right)\right], \label{eq:12}
\eea
where $\ts=(p_1-k_1)^2-\ms^2$, and $\us=(p_1-k_2)^2-\ms^2$ are again
mass-subtracted Lorentz-invariant Mandelstam variables, $p_i^\mu$ are now the
four-momenta of the incoming partons, and $N_C=3$ denotes the number of colors.
The unpolarized double-resolved photon cross section is then given by
\beq
 \sigma^{\rm res}_{\gamma\gamma} =
 \sum_{i,j=q,\bar{q},g}\int\d x_1 f_{i/\gamma}(x_1,M^2)
 \d x_2 f_{j/\gamma} (x_2,M^2) {1\over2s}{\d\tg\over8\pi s}{1\over2}
 \overline{|{\cal M}_{ij}|}^2,
\eeq
where $s=x_1x_ss_{\gamma\gamma}$ is the partonic center-of-mass energy,
$x_{1,2}$ are the longitudinal momentum fractions of the partons $i$ and $j$
in the photons, and $M$ is the factorization scale, which we identify with
the gluino mass. The variation of the double-resolved photon cross section
with this scale, which amounts to $\pm 35$ \% in leading order for a
variation of $M$ about $\mg$ by a factor of four, will be considerably
reduced, once the corresponding higher-order direct processes are included.

In contrast to the gluon-gluon initiated matrix element in Eq.\ (\ref{eq:12}),
which involves only gluons and gluinos and depends only on the gluino mass, the
quark-antiquark initiated matrix element in Eq.\ (\ref{eq:11}) involves also
quarks and squarks and thus depends also on the masses of these (s)particles.
Since we are working in the collinear limit at high center-of-mass energies,
we neglect the masses of the five light initial quark and antiquark flavors
and take their momentum distributions inside the photon, $f_{i/\gamma}(x,M^2)$,
from a leading-order, five-flavor fit to the photon structure function
\cite{Gluck:1991jc}. For center-of-mass energies $\sqrt{s}_{\gamma\gamma}>2
(m_t+\mg)>748.6$ GeV (relevant for $\mg>200$ GeV), top quarks can
contribute through the process $\gamma\gamma\to t\bar{t}\go\go$ at
$\O(\alpha^2\alpha_s^2)$. While this process is of the same order as the 
direct contributions considered in the previous Section, it is phase-space
suppressed by the larger mass of the final state and will therefore be
neglected.

\begin{figure}
 \centering
 \epsfig{file=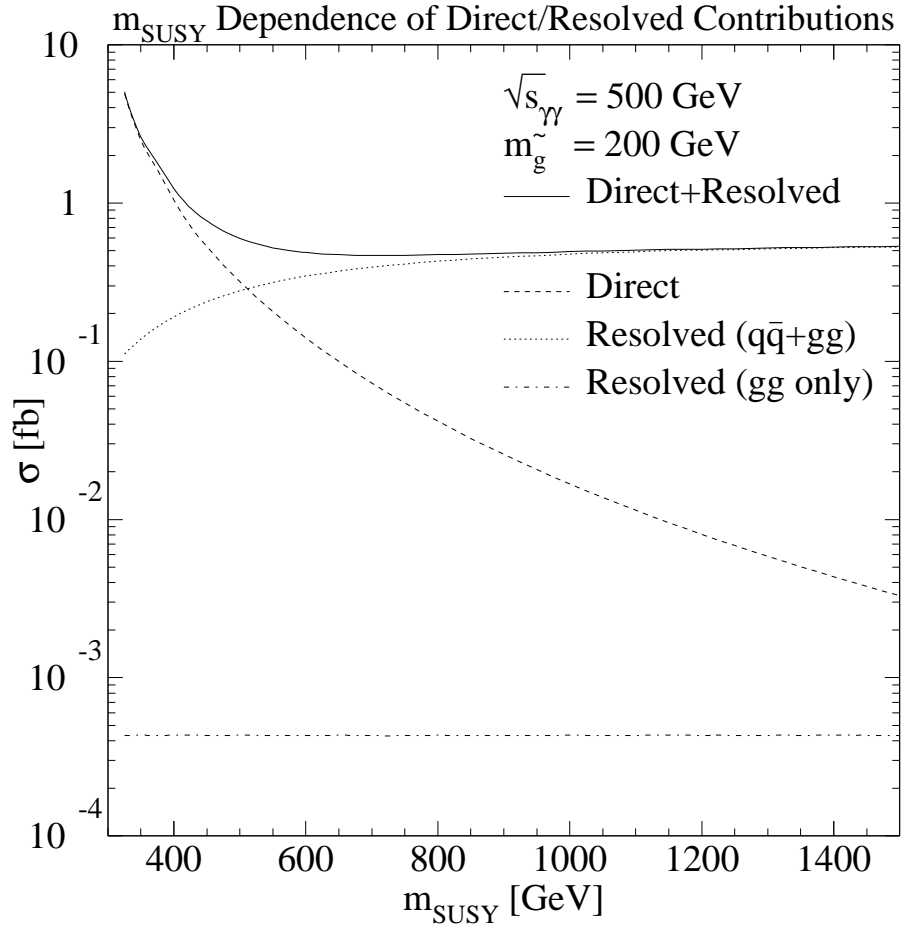,width=0.8\columnwidth}
 \caption{\label{fig:14}Dependence of the direct, resolved, and $gg$ initial
 state contributions to the gluino pair production cross section in unpolarized
 photon-photon collisions on the squark mass parameter $\msusy$.}
\end{figure}

In Fig.\ \ref{fig:14}, the direct contribution (dashed curve) falls steeply
%
%
with the squark mass parameter $\msusy$, while the small resolved gluon-gluon
initiated channel (dot-dashed curve) is, of course, independent of the squark
mass. Interestingly, the resolved quark-antiquark initiated contribution,
which coincides with the total resolved cross section (dotted curve),
increases with $\msusy$. The reason is that the $s$- and $t$-channel diagrams
in Fig.\ \ref{fig:9} interefere destructively, and the interference
contributions to Eq.\ (\ref{eq:11}) decrease in magnitude as $\msusy$
increases. As a consequence, the total gluino pair production cross section
(full curve) remains large even for large squark masses and becomes independent
of $\msusy$ already for moderate squark masses.

\begin{figure}
 \centering
 \epsfig{file=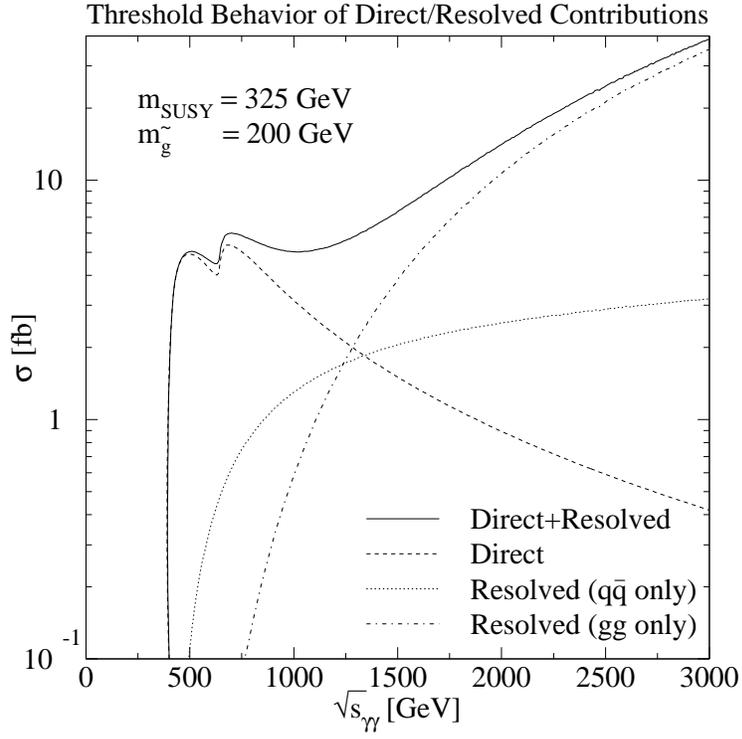,width=0.63\columnwidth}
 \epsfig{file=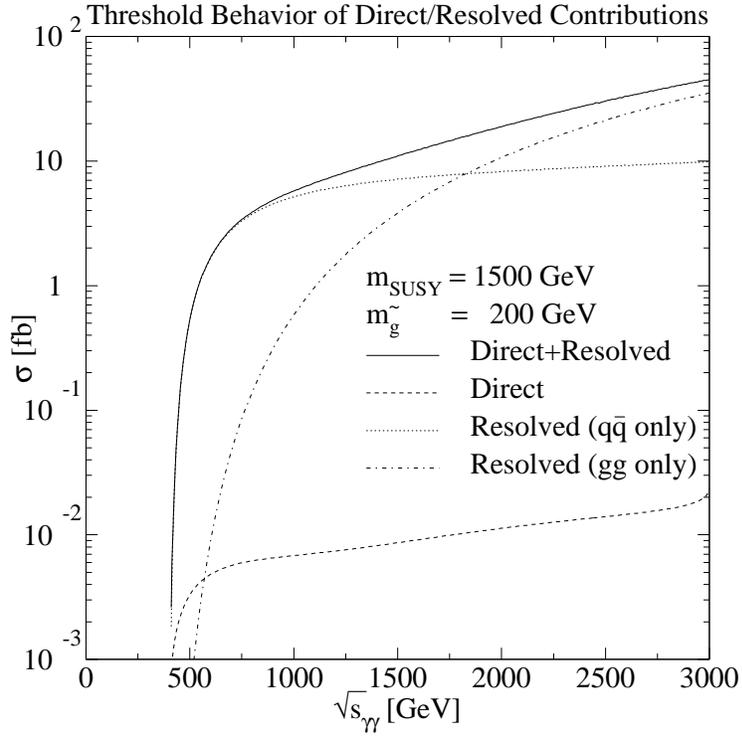,width=0.63\columnwidth}
 \caption{\label{fig:13}Threshold behavior of the direct, $q\bar{q}$, and $gg$
 initial state contributions to the gluino pair production cross section in
 unpolarized photon-photon collisions for a small (upper plot) and large
 (lower plot) squark mass parameter $\msusy$.}
\end{figure}

Figure \ref{fig:13} shows the threshold behavior of the direct (dashed curves),
%
%
quark-antiquark (dotted curves), and gluon-gluon initiated cross sections
(dot-dashed curves) for small (top) and large (bottom) values of $\msusy$.
In the first scenario, the direct channel dominates at threshold as expected,
while in the second case it is completely negligible in the entire
center-of-mass energy range. Instead, the threshold behavior of the total cross
section is dominated by the quark-antiquark channel, and the gluon-gluon
process contributes significantly at larger values of $\sqrt{s}_{\gamma
\gamma}$.

Resolved contributions are therefore only important (1) if $\sqrt{s}_{\gamma
\gamma}\gg2\mg$, {\it i.e.} far above the gluino pair production threshold, or
(2) if $\ms\gg\mg$. The first case is of little practical importance. For
$\ms\leq\mg$, it would in fact be desirable to suppress the resolved
contributions so that the photon-photon center-of-mass energy is fully
available in threshold scans. This should be feasible by reconstructing the
observed momentum fractions of the partons in the photon from the gluino pair
and requiring them to be large. The second case is more interesting and may
allow for a gluino mass determination in scenarios that would be unobservable
in $e^+e^-$ or direct $\gamma\gamma$ collisions.

\section{Polarized Electron-Electron Scattering}
\label{sec:4}

In practice, high-energy photon collisions will become feasible through the
backscattering of laser photons from electron beams at future linear colliders.
The unpolarized and polarized laser backscattering spectra are
\cite{Ginzburg:1982yr}
\bea
 f_{\gamma/e}(x) &=& \frac{1}{N_c+2 \lambda_e P_c N_c'} \left\{1-x
 +\frac{1}{1-x}-\frac{4 x}{X (1-x)}+\frac{4 x^2}{X^2 (1-x)^2} \right.\\
 && \hspace*{28mm} \left. - 2 \lambda_e P_c \frac{x (2-x) [x (X+2)-X]}
 {X (1-x)^2} \right\} \nonumber
\eea
and
\bea
 \Delta f_{\gamma/e}(x) &=& \frac{1}{N_c+2 \lambda_e P_c N_c'}
 \left\{ 2 \lambda_e \frac{x}{1-x}
 \left[ 1+(1-x) \left( 1-\frac{2 x}{(1-x) X}\right)^2\right]\right.\\
 &&\hspace*{28mm}\left.+ P_c \left( 1-\frac{2 x}{(1-x) X}\right)
 \left(1-x+\frac{1}{1-x}\right) \right\}, \nonumber
\eea
where
\beq
 N_c = \left[1-\frac{4}{X}-\frac{8}{X^2}\right] \ln(1+X)+\frac{1}{2}
 + \frac{8}{X}-\frac{1}{2 (1+X)^2} \\
\eeq
and
\beq
 N_c'= \left[\left(1+\frac{2}{X}\right)\ln(1+X)-\frac{5}{2}+\frac{1}{1+X}
 -\frac{1}{2(1+X)^2}\right]
\eeq
are related to the total Compton cross section and $\xi_i=\Delta f_{\gamma/e}
(x_i)/f_{\gamma/e}(x_i)$ is the mean helicity of the backscattered photon
$i$.

The photon spectra depend on the center-of-mass energy of the electron-laser
photon collision $s_{e\gamma}$ through the parameter $X=s_{e\gamma}/m_e^2-1$,
whose optimal value is determined by the threshold for $e^+e^-$ pair
production. In the strong fields of the laser waves, the electrons (or
high-energy photons) can interact simultaneously with several laser photons,
so that the optimal value of $X$ increases from $(2+\sqrt{8})$ to $(2+\sqrt{8})
(1+\xi^2)\simeq6.5$ \cite{Burkhardt:2002vh}. If the parameter $X$ is kept
fixed, the laser backscattering spectra become independent of the electron
beam energy. A large fraction of the photons is then produced close to the
kinematic limit $x<x_{\max}=X/(X+1)=0.8\bar{6}$. The monochromaticity of the
produced photons can be improved further by choosing the helicity of the 100\%
polarized laser photons opposite to that of the 80\% polarized electrons.

Since the low-energy tail of the photon spectrum is neither useful nor well
understood, we use only the high-energy peak with $x>0.8\, x_{\max}=
0.69\bar{3}$ and normalize our cross sections such that the expected
number of events can be obtained through simple multiplication with the
envisaged photon-photon luminosity of 100-200 fb$^{-1}$/year. This requires
reconstruction of the total final-state energy, which may be difficult due to
the missing energy carried away by the (typically two) escaping lightest SUSY
particles (LSPs). However, high-energy photon collisions allow for cuts on the
relative longitudinal energy in addition to the missing-$E_T$ plus multi-jet,
top or bottom quark, and/or like-sign lepton analyses performed at hadron
colliders, and sufficiently long-lived gluinos can be identified by their
typical $R$-hadron signatures.

The total cross section for gluino pair production in polarized
electron-electron collisions is given by
\bea
 \sigma_{ee} &=& \int \d x_1 f_{\gamma/e}(x_1) \d x_2 f_{\gamma/e}(x_2) \d\tg\\
 &\times&
 \le{1\over2}(1+\xi_1)(1+\xi_2)\frac{\d\sigma_{\rm dir}^{++}}{\d\tg}
   +{1\over2}(1+\xi_1)(1-\xi_2)\frac{\d\sigma_{\rm dir}^{+-}}{\d\tg}\rp
    \nonumber\\ &+&\hspace*{1.6mm}\lp
    {1\over2}(1-\xi_1)(1+\xi_2)\frac{\d\sigma_{\rm dir}^{-+}}{\d\tg}
   +{1\over2}(1-\xi_1)(1-\xi_2)\frac{\d\sigma_{\rm dir}^{--}}{\d\tg}\re,
 \nonumber
\eea
where parity conservation in (SUSY-)QCD guarantees that $\sigma^{++}_{\rm dir}=
\sigma^{--}_{\rm dir}$ and $\sigma^{+-}_{\rm dir}=\sigma^{-+}_{\rm dir}$.
For resolved processes, the photon densities are replaced by parton densities
in the electron,
\beq
 f_{a/e}(x,M^2)  =  \int_x^1 \frac{\d y}{y} \hspace*{3mm} f_{\gamma/e}
 \left(\frac{x}{y}\right) \hspace*{3mm} f_{a/\gamma} (y,M^2)
\eeq
and
\beq
 \Delta f_{a/e}(x,M^2)  =  \int_x^1 \frac{\d y}{y} \Delta f_{\gamma/e}
 \left(\frac{x}{y}\right) \Delta f_{a/\gamma} (y,M^2).
\eeq

\begin{figure}
 \centering
 \epsfig{file=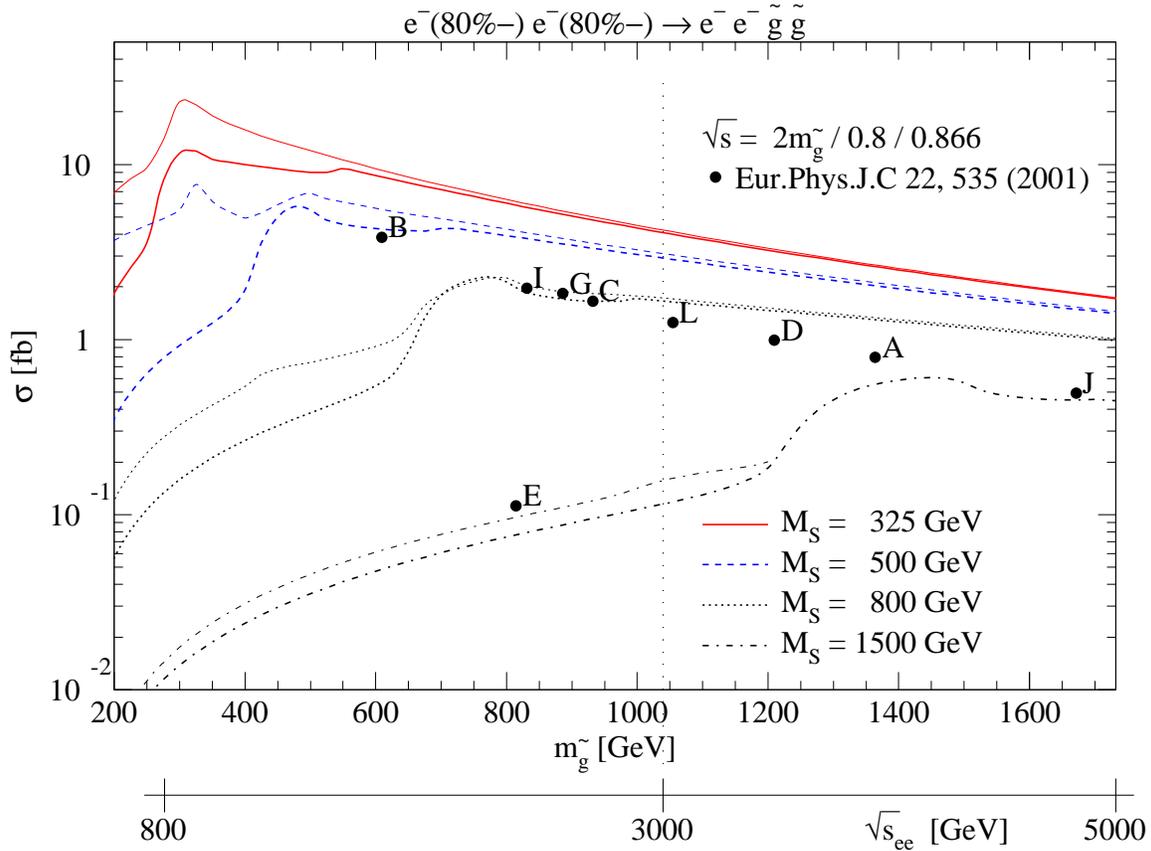,width=\columnwidth}
 \caption{\label{fig:3}Dependence of the gluino pair production cross section
 in $\gamma\gamma$ collisions on the universal squark mass $\msusy$ for
 no squark mixing (thick curves) and maximal top-squark mixing (thin curves).
 Also shown are the cross sections expected for the post-LEP SUSY
 benchmark points in Ref.\ \cite{Battaglia:2001zp} (full points).
 The photon-photon luminosity has been normalized to unity in the high-energy
 peak.}
\end{figure}

\begin{figure}
 \centering
 \epsfig{file=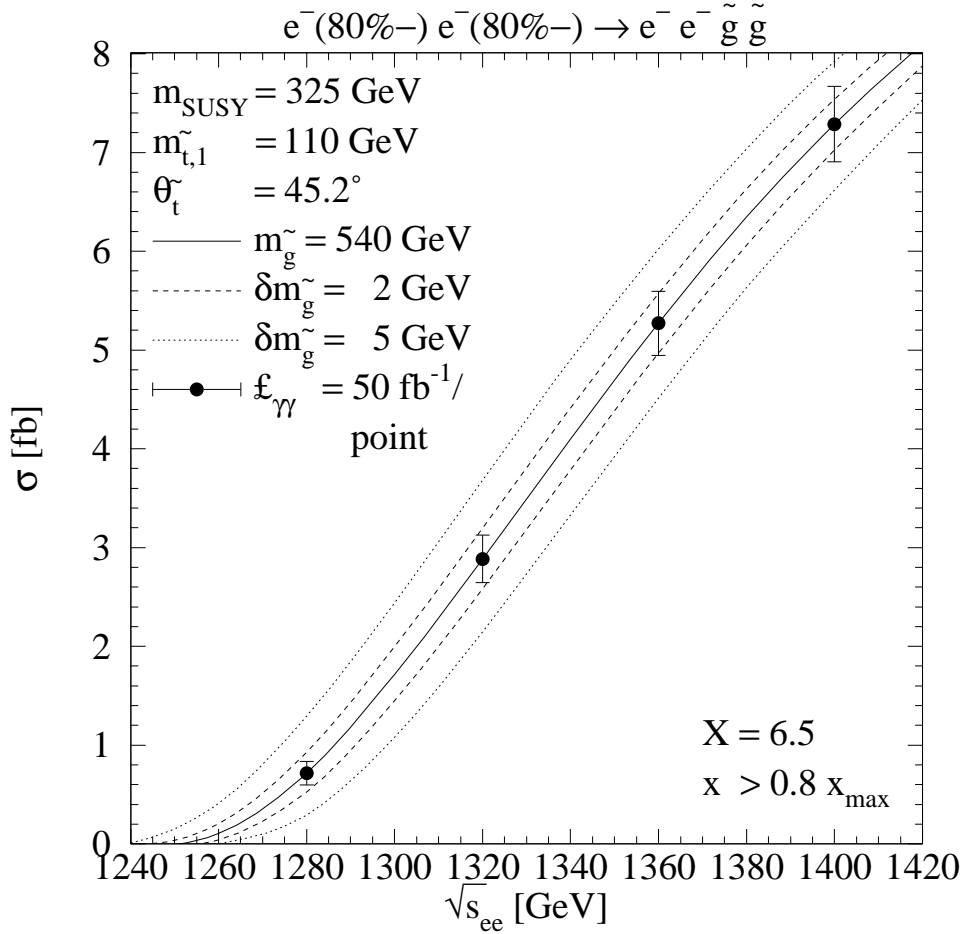,width=0.8\columnwidth}
 \caption{\label{fig:2}Sensitivity of the gluino pair production cross section
 in polarized direct photon collisions to the mass of the pair-produced gluino.
 The photon-photon luminosity has been normalized to unity in the high-energy
 peak.}
\end{figure}

In Fig.\ \ref{fig:3} we show that gluino pair production in direct
%
%
$\gamma\gamma$ collisions decreases with the universal squark mass $\msusy$,
but depends only weakly on the top-squark mixing. This is in sharp contrast to
the results obtained in $e^+e^-$ annihilation \cite{Berge:2002ev,Berge:2002xc}.
In this plot, the $e^-e^-$ center-of-mass energy is chosen close to the
threshold for gluino pair production ($\sqrt{s}=2\mg/0.8/0.8\bar{6}$)
and is varied simultaneously with $\mg$.
Also shown in Fig.\ \ref{fig:3} are several post-LEP SUSY benchmark points,
which have recently been proposed within the framework of the constrained MSSM
\cite{Battaglia:2001zp}. Studies similar to those performed in Fig.\
\ref{fig:2} show that with the exception of point E, where only about ten
events per year are to be expected, the gluino mass can be dermined with a
precision of $\pm20$ GeV (point J) or better.

\begin{figure}
 \centering
 \epsfig{file=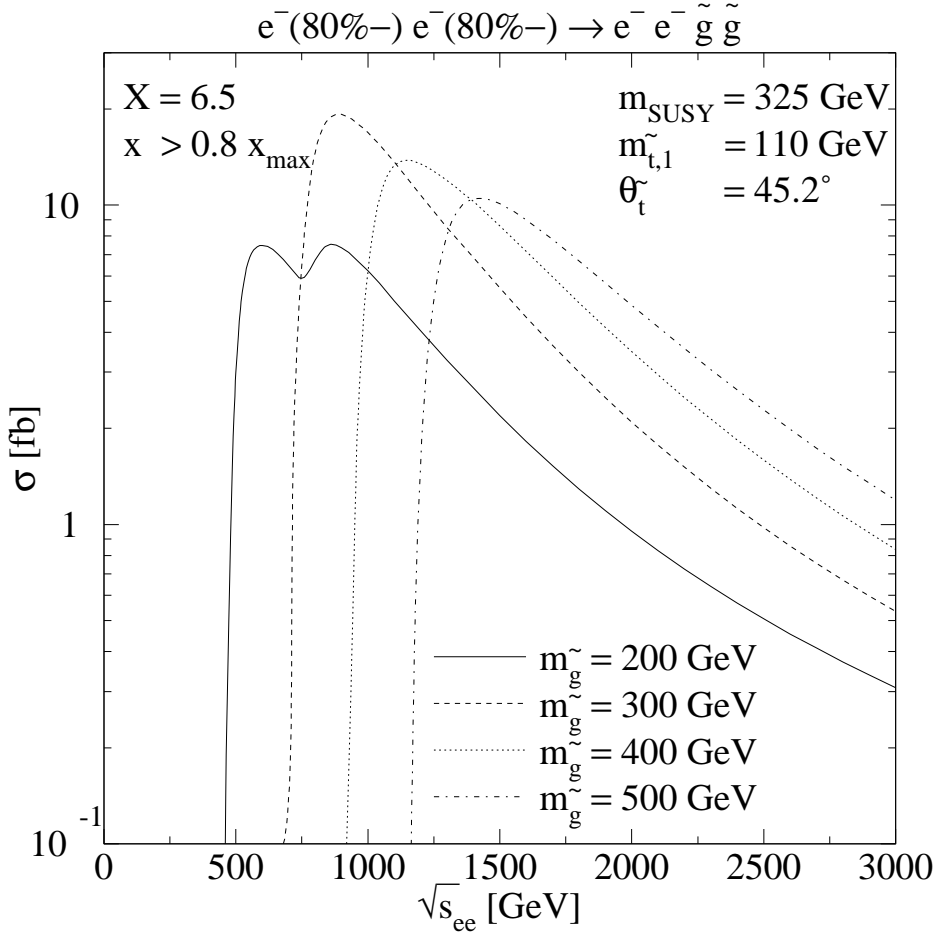,width=0.8\columnwidth}
 \caption{\label{fig:1}The gluino pair production cross section in polarized
 direct photon collisions as a function of the $e^-e^-$ center-of-mass energy
 for various gluino masses. The photon-photon luminosity has been normalized
 to unity in the high-energy peak.}
\end{figure}

For gluino masses between 200 and 500 GeV, the total cross section for gluino
pair production in polarized electron-electron collisions with
laser-backscattered photons is shown in Fig.\ \ref{fig:1} as a function
%
%
of the electron-electron center-of-mass energy $\sqrt{s}_{ee}$. The
corresponding $e^+e^-$ annihilation cross section stays below 0.1 fb and falls
steeply with $\mg$, so that gluino pair production will be unobservable for
$\mg>500$ GeV irrespective of the collider energy \cite{Berge:2002ev,
Berge:2002xc}. In contrast, the $\gamma\gamma$ cross section reaches several
fb for a wide range of $\mg$. In $e^+e^-$ annihilation the gluinos are
produced as a $P$-wave and the cross section rises rather slowly, whereas in
$\gamma\gamma$ collisions they can be produced as an $S$-wave and the cross
section rises much faster. This is particularly true for identical helicities
of the initial photons (see Fig.\ \ref{fig:11}).

The steep threshold behavior can be observed even more clearly in Fig.\
\ref{fig:2}, where the sensitivity of a photon collider to the gluino mass
%
%
is investigated. For the CERN LHC experiments, a precision of $\pm30\,...\,60$
($12\,...\,25$) GeV is expected for gluino masses of 540 (1004) GeV
\cite{:1999fr,Abdullin:1998pm}. If the masses and mixing angle(s) of the top
(and bottom) squarks are known, a statistical precision of $\pm5\,...\,10$ GeV
can be achieved in $e^+e^-$ annihilation for $\mg=200$ GeV for an integrated
luminosity of 100 fb$^{-1}$ per center-of-mass energy point \cite{Berge:2002ev,
Berge:2002xc}. A precision of $\pm2\,...\,5$ GeV may be obtained at a TeV-scale
photon collider for $\mg=540$ GeV and an integrated photon-photon luminosity of
50 fb$^{-1}$ per point, provided that the total final-state energy can be
sufficiently well reconstructed. Of course, uncertainties from a realistic
photon spectrum and the detector simulation add to the statistical error.

\begin{figure}
 \centering
 \epsfig{file=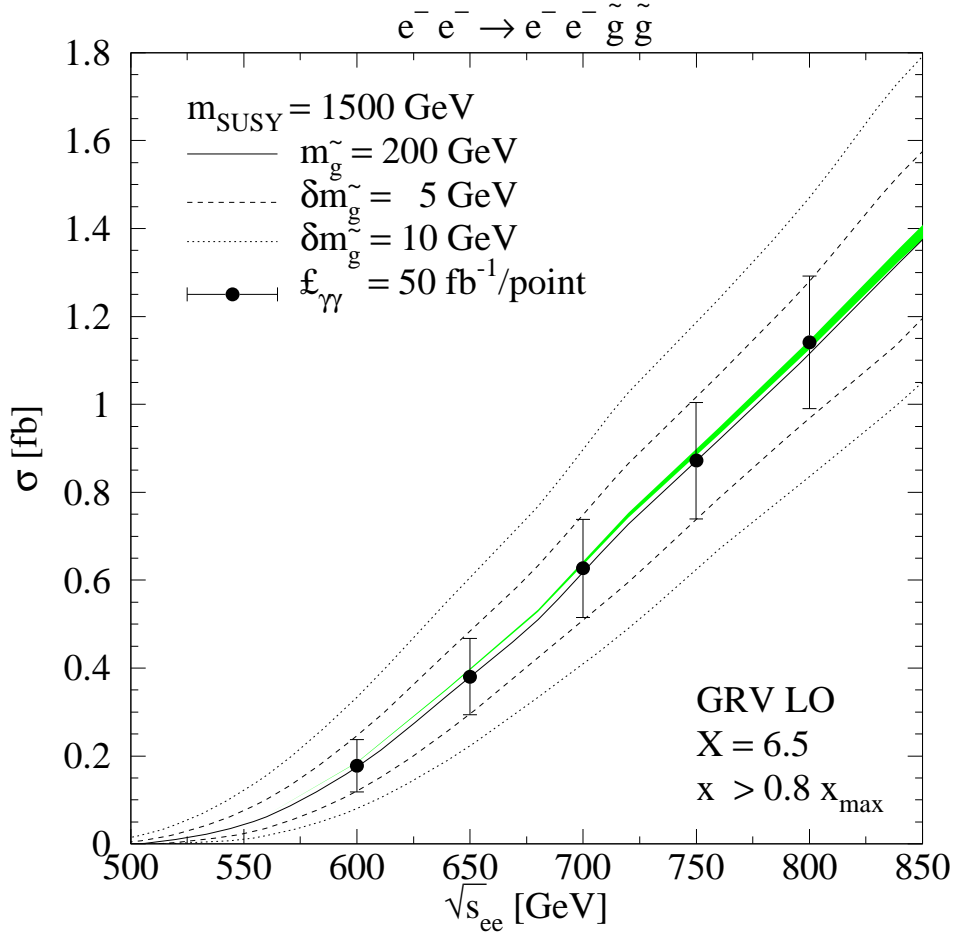,width=0.8\columnwidth}
 \caption{\label{fig:16}Sensitivity of the gluino pair production cross section
 in unpolarized resolved photon collisions to the mass of the pair-produced
 gluino. The photon-photon luminosity has been normalized to unity in the
 high-energy peak.}
\end{figure}

In Fig.\ \ref{fig:16} we demonstrate that even in scenarios with a very large
%
%
squark mass parameter $\msusy=1500$ GeV and no squark mixing a gluino mass
of $\mg=200$ GeV could be determined with a precision of $\pm5\,...\,10$ GeV.
The cross section is then dominated by the quark-initiated
resolved contributions (see the lower plot in Fig.\ \ref{fig:13}). The shaded
band indicates the uncertainty from a variation of parton densities in the
photon from GRV LO \cite{Gluck:1991jc} to SaS 1D and 1M \cite{Schuler:1995fk}.
It turns out to be fairly small, since the quark densities at large $x$ are
well constrained from photon structure function data \cite{Klasen:2002xb}.

\section{Conclusions}
\label{sec:5}

In conclusion, the reconstruction of the SUSY Lagrangian and the precise
determination of its free parameters are among the paramount objectives of any
future linear $e^+e^-$ collider. Determination of the gluino mass and coupling
will, however, be difficult, since the gluino couples only strongly and its
pair production cross section in $e^+e^-$ annihilation suffers from large
cancellations in the triangular quark/squark loop diagrams. A photon collider
may therefore be the only way to obtain precise gluino mass determinations and
visible gluino pair production cross sections for general squark masses and
would thus strongly complement the physics program feasible in $e^+e^-$
annihilation. For recently proposed typical post-LEP SUSY benchmark points,
gluino pairs can be produced from laser-backscattered photons with cross
sections between 0.1 fb and 4 fb, so that a gluino mass of 540 (1670) GeV
may be determined with a precision of $\pm2...5$ ($\pm 20$) GeV  or better.


\begin{acknowledgments}

We thank A.\ de Roeck for bringing Ref.\ \cite{Battaglia:2001zp} to our
attention and B.A.\ Kniehl and G.\ Kramer for a careful reading of the
manuscript. This work has been supported by Deutsche Forschungsgemeinschaft
through Grant No.\ KL~1266/1-3 and through Graduiertenkolleg {\it Zuk\"unftige
Entwicklungen in der Teilchenphysik}.

\end{acknowledgments}



\end{document}